\documentclass[
reprint,
superscriptaddress,
bibnotes,
 amsmath,amssymb,
pra,
floatfix,
]{revtex4-1}

\usepackage{graphicx}
\usepackage{dcolumn}
\usepackage{bm}

\begin{document}

\preprint{APS/123-QED}

\title{Direct measurement of the orbital angular momentum mean and variance in an arbitrary paraxial optical field}

\author{Bruno Piccirillo}
\email{bruno.piccirillo@na.infn.it}
\affiliation{%
Dipartimento di Fisica, Universit\`a di Napoli Federico II, Compl. Univ. di Monte S. Angelo, Napoli, Italy}%
\author{Sergei Slussarenko}%
\affiliation{%
Dipartimento di Fisica, Universit\`a di Napoli Federico II, Compl. Univ. di Monte S. Angelo, Napoli, Italy}%
\author{Lorenzo Marrucci}
\affiliation{%
Dipartimento di Fisica, Universit\`a di Napoli Federico II, Compl. Univ. di Monte S. Angelo, Napoli, Italy}%
\affiliation{%
CNR-SPIN, Compl. Univ. di Monte S. Angelo, Napoli, Italy
}%
\author{Enrico Santamato}
\affiliation{%
Dipartimento di Fisica, Universit\`a di Napoli Federico II, Compl. Univ. di Monte S. Angelo, Napoli, Italy}%


\date{\today}

\begin{abstract}
We introduce and experimentally demonstrate a method for measuring at the same time the mean and the variance of the photonic orbital angular momentum (OAM) distribution in any paraxial optical field, without passing through the acquisition of its entire angular momentum spectrum. This method hence enables one to reduce the infinitely many output ports required in principle to perform a full OAM spectrum analysis to just two. The mean OAM, in turn, provides direct access to the average mechanical torque that the optical field in any light beam is expected to exert on matter, for example in the case of absorption. Our scheme could also be exploited to weaken the strict alignment requirements usually imposed for OAM-based free-space communication.
\begin{description}
\item[PACS numbers]
42.50.Tx, 42.25.-p, 42.87.Bg
\end{description}
\end{abstract}

\pacs{Valid PACS appear here}
\maketitle


Nowadays, the most relevant properties of the optical radiation can be measured with great accuracy, and essentially all features of light have been used to encode and transfer data or to realize optical devices. Particular interest, among the properties of light, has been recently gained by the angular momentum and the ensuing rotational effects induced in matter, e.g.\ by the circular polarization of light or by the optical ray-torsion \cite{allen99a,piccirillo04b}. These two cases correspond to the natural subdivision of the optical angular momentum of a (paraxial) light beam into a spin part (SAM), associated with the polarization degree of freedom, and an orbital part (OAM), associated with the phase front structure \cite{allen99a,allen92,vanenk94b}. In the last two decades, the OAM technology has advanced significantly \cite{yao11,marrucci11}, but it is still lacking important tools. In this Letter, we tackle in particular the problem of measuring the mean of the orbital angular momentum of a paraxial light beam, together with its variance. The photon OAM along the propagation axis $z$ can be defined as the quantum average of the operator $\hat{L}_z=x\hat{p}_y - y\hat{p}_x$ in a given state of radiation, where $\hat{\mathbf{p}}=-i\hbar\nabla$ is the photon momentum operator and $\hbar$ is the reduced Planck constant \cite{vanenk94b}. Consistently, the fraction of the total intensity belonging to the helical transverse modes characterized by the phase factor ${\rm exp}(i l \phi)$, where $\phi$ is the azimuthal coordinate around the $z$ axis, can be identified with the probability of obtaining $\hbar l$ when measuring the OAM of the photon in that state \cite{allen92,vanenk92}. The measurement of the optical OAM is a long-standing and challenging problem, both in quantum and classical optics. Most existing methods are actually aimed at measuring either the topological charge of the field or the power spectrum of the helical modes it contains. The topological charge can be obtained from inspection of suitable interference or diffraction patterns, e.g.\ using a multi-pinhole interferometer \cite{berkhout08} or triangular apertures \cite{hickmann10}. However, the topological charge of a wave is not uniquely linked to its OAM content, this simple identification being valid only for OAM eigenmodes. The spectrum of helical eigenmodes of a given optical field, sometimes named ``spiral spectrum'', can be obtained by filtering or splitting the wave according to the OAM eigenmode, i.e.\ using mode selectors or mode sorters. Typical OAM eigenmode selectors are for example based on spiral phase plates or pitchfork holograms in combination with single-mode fibers \cite{beijersbergen94,mair01}, or more recently on the so-called ``$q$-plates'' in combination with suitable polarization optics \cite{marrucci06,karimi09a}. However, to obtain a complete OAM spectrum, the parallel or sequential detection of many different OAM modes (ideally, an infinite number) is required. This can be for example achieved using special holograms, although with an efficiency that scales inversely with the number of sorted modes~\cite{khonina00,gibson04}. A highly efficient scheme, although with strong practical limitations, can be based on a cascaded sequence of interferometers \cite{leach02,wei03}. A more convenient efficient OAM mode sorter was demonstrated recently, using a pair of custom refractive elements to ``unfold'' the azimuthal phase of OAM modes \cite{berkhout10,berkhout11,lavery12}. Finally, an alternative indirect method to determine the OAM spectrum is based on measuring the spatial-resolved second-order field correlations, e.g.\ using a multi-pinhole interferometer as in Ref.~\cite{berkhout09}. The full OAM spectral analysis of these methods has the merit of returning the whole probability distribution of OAM and, hence, all the statistical moments of that observable, including the mean and the variance. However, given the considerable overhead arising from the need of measuring a large (ideally infinite) number of probabilities, it is clear that a technique to measure directly the mean OAM, and eventually its variance, without passing through the determination of the OAM spectrum, could be very useful. Such quantities, although generally inadequate for fully reconstructing the distribution, may provide important information about the source of the radiation field and about the OAM conservation in radiation-matter interactions. The mean OAM is also immediately related to the mean mechanical torque exerted by light on a medium \cite{padgett97,santamato02}.

Here, we introduce and experimentally demonstrate a method to measure directly the first two moments of the OAM distribution, i.e.\ the expectation values of the observables $L_z$ and $L_z^2$, of an arbitrary paraxial light beam, while avoiding the full spectral analysis. The variance of the OAM can then of course be obtained as $(\Delta L_z)^2=\langle L_z^2\rangle-\langle L_z\rangle^2$. Our approach is based on a polarization homodyne detection scheme, working both in continuous-wave (cw) and in photon-counting regimes, reminiscent of that theoretically proposed in Ref.~\cite{zambrini06} for measuring the whole OAM spectrum. We then validate our method with a series of tests on controlled input beams of varying OAM distributions.

The principle introduced here to directly measure $\langle L_z \rangle$ could be advantageously exploited to weaken the restrictions on the sender-receiver alignment typical of free-space communication systems. It is well-known, in fact, that $\langle L_z \rangle$ is invariant under translations in the $xy$ plane, i.e.\ orthogonal to the propagation direction of the field.

\begin{figure}
\includegraphics[scale=0.06]{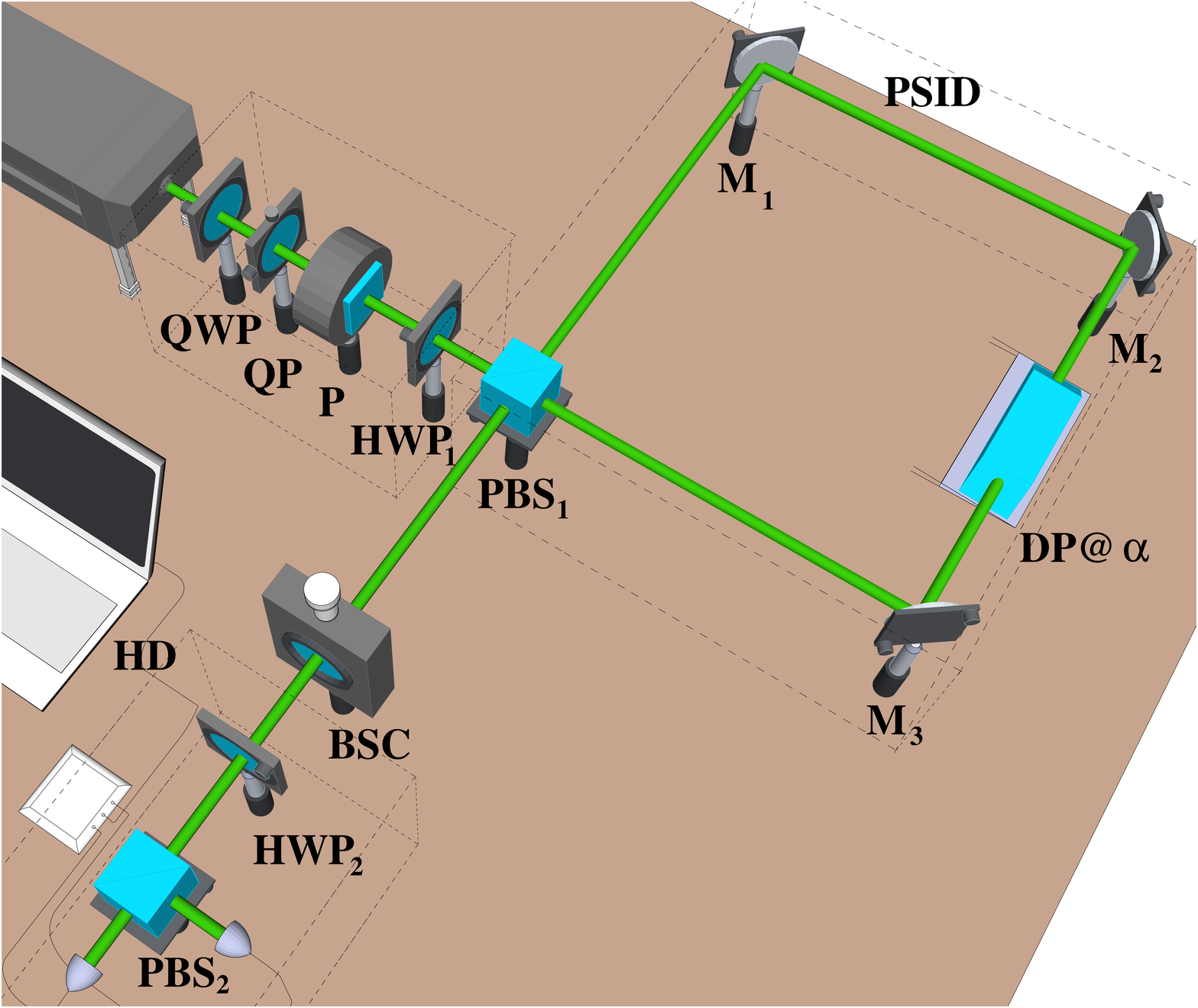}
\caption{\label{fig:fig1} Schematic of the setup for measuring the first and second moments of the OAM probability distribution carried by a paraxial light beam. An initially Gaussian beam is converted into an OAM superposition state using a quarter-waveplate QWP and a $q$-plate QP. A polarizer P and a half-waveplate HWP$_1$ are then used to prepare the polarization in a linear state oriented at 45$^\circ$ with respect to the axes of the polarizing beam splitter PBS$_1$. This PBS$_1$ is the input/output port of a polarizing Sagnac interferometer, whose path is closed by mirrors M$_1$, M$_2$, M$_3$, that contains the Dove prism (PSID) used to rotate the two counterpropagating beams with respect to each other. At the output of the PSID, a Babinet-Soleil compensator (BSC) is used to adjust the relative phase-shift $\delta$. The last stage is the balanced polarizing homodyne detector HD, with the axes rotated by 45$^\circ$ with respect to the PSID axes through the half-waveplate HWP$_2$.}
\end{figure}

\textit{Concept of the method}. Our method is essentially based on preparing first two copies of the input beam that are rotated with respect to each other by a small angle around the propagation axis, and next letting them interfere in a polarization homodyne detector. The two rotated copies are prepared by using a Sagnac Polarizing Interferometer containing a Dove prism (PSID) \cite{slussarenko10}. The details of the light manipulation are illustrated in Fig.~\ref{fig:fig1}. The base of the Dove prism is tilted by the angle $\alpha$ with respect to the PSID plane. Consequently, the beams propagating in opposite directions along the arms of the PSID suffer opposite azimuthal angular shifts of $2\alpha$, leading to their relative rotation of $4\alpha$. After exiting the PSID, these two ortogonally polarized beams are superimposed in the final homodyne detection stage, as shown in Fig.\ \ref{fig:fig1}.
In detail, let $\bm{E}=E(r,\phi,z){\bm a}$ be the input optical field, where ${\bm a}=(\bm{x}+\bm{y})/\sqrt{2}$, $\bm x$ ($\bm y$) being the unit vector along the $x$ ($y$) axis. The output field after the PSID and the Babinet-Soleil compensator used to adjust the phase difference $\delta$ is $E_+{\rm e}^{{\rm i}\delta/2}\bm {x}/\sqrt{2}+E_-{\rm e}^{{-\rm i}\delta/2}\bm {y}/\sqrt{2}$, where $E_\pm = E(r,\phi \pm 2\alpha,z)$. The field projections along the antidiagonal and diagonal polarization directions ${\bm a}=(\bm{x}+\bm{y})/\sqrt{2}$ and ${\bm d}=(\bm{x}-\bm{y})/\sqrt{2}$, respectively, are then given by
\begin{eqnarray}
{\bm E}_a&=&\left(E_+ {\rm e}^{{\rm i}\delta/2}+E_-{\rm e}^{{-\rm i}\delta/2}\right){\bm a}/\sqrt{2} \label{eq:homodyne1}\\
{\bm E}_d&=&\left(E_+ {\rm e}^{{\rm i}\delta/2}-E_-{\rm e}^{{-\rm i}\delta/2}\right){\bm d}/\sqrt{2} \label{eq:homodyne2}.
\end{eqnarray}
The interference fields ${\bm E}_a$ and ${\bm E}_d$ are collected by two inter-calibrated photodetectors and electronically processed to provide the sum and difference homodyne signals
\begin{eqnarray}\label{eq:sum_and_diff}
P_0&=& P_a + P_d\\
\Delta P &=& P_a - P_d =\sum_l {P_0^l \cos{(4 l \alpha + \delta})},
\end{eqnarray}
where $P_a\propto|E_a|^2$ and $P_d\propto|E_d|^2$ are the powers of the interference fields (\ref{eq:homodyne1}) and (\ref{eq:homodyne2}), $P_0$ is the total power of the output beam and $P_0^l$ the power content of the $l$-mode in the OAM expansion. In the limit of small $\alpha$,
\begin{equation}\label{eq:expansion}
\frac{\Delta P}{P_0}=\cos{\delta} -4 \alpha \sin{\delta} \langle L_z \rangle -8\alpha^2 \cos{\delta}\langle L_z^2 \rangle + O(\alpha^3).
\end{equation}
Assuming $\alpha$ to be small enough to allow truncation of the signal expansion to 2$^{\rm nd}$ order, we obtain $\langle L_z \rangle = -(1/4\alpha) \Delta P /P_0$ for $\delta=(2k+1)\pi/2$, and $\langle L_z^2 \rangle = (1/8\alpha^2) (1-\Delta P /P_0)$ for $\delta = 2 k \pi$, with $k$ being an integer. The preselected value of $\alpha$ is set through the goniometer $G$ ($\pm 0.008\ensuremath{^\circ}$) and an accurate estimate for its value was then obtained by the calibration procedure. Zeroing and calibration of the apparatus were carried out by adopting as standard input the OAM imparted by perfectly tuned $q$-plates~\cite{marrucci06,marrucci11}, for several values of $q$ (with $q\leq 8$), onto the high-quality TEM$_{00}$ mode generated by a single-mode cw laser (wavelength $\lambda=532$~nm). The selection criterium for $\alpha$ is a critical point of the method. In fact, it sets an upper limit to the maximum value of $l$ that can be measured within a specified accuracy (upper cutoff) and controls the signal-to-noise ratio. For instance, to set the cutoff $l_{max}=100$, maintaining within $1\%$ the theoretical accuracy in the measurement of both $\langle L_z \rangle$ and $\langle L_z^2 \rangle$, an $\alpha\approx 0.05\ensuremath{^\circ}$ is to be selected. Since, the signal returning $\langle L_z \rangle$ is proportional to $\alpha$ while that returning $\langle L_z^2 \rangle$ is proportional to $\alpha^2$, to accurately discern the eigenstate $l=1$, the noise in the signals must be lower than $\approx 10^{-5}$. This requires a very precise control of the optical path within PSID, which is reflected into a good control of the noise in the phase retardation $\delta$. Our experiment was carried out with $\alpha=(0.85\pm0.01)^{\circ}$, corresponding to $l_{max}=10$ and a maximum acceptable noise in the signals of $\approx 10^{-3}$.
\begin{figure}
\includegraphics[scale=0.9]{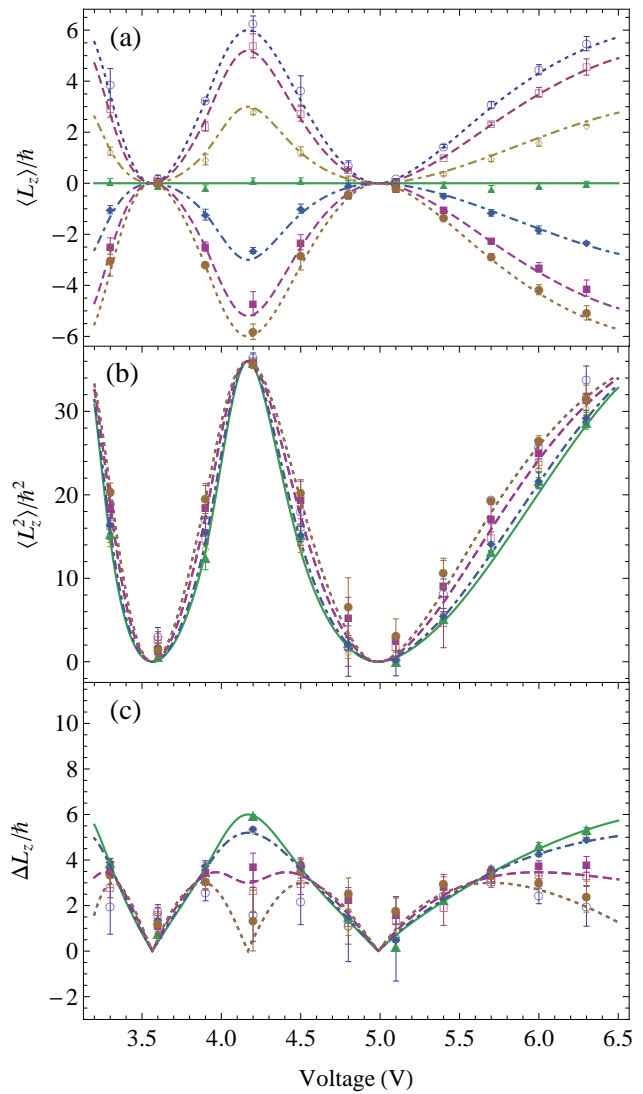}
\caption{\label{fig:fig2} Behavior of the first (a) and second (b) moment and of the root mean square deviation (c) of the OAM distribution of the beam prepared using a q-plate, as a function of the q-plate tuning voltage controlling its retardation $\theta$ and for different values of the polarization helicity $s_3$ before the q-plate. The two parameters $\theta$ and $s_3$, together with the q-plate charge $q$, define the specific OAM superposition of states having $l=0$ and $l=\pm2q=\pm6$. Experimental data points for $s_3=1.0 \; (-1.0)$ are represented by $\circ$ ($\bullet$), for $s_3=0.87 (-0.87)$ by $\square$ ($\blacksquare$), for $s_3=0.50 (-0.50)$ by $\lozenge$ ($\blacklozenge$) and for $s_3=0$ by $\blacktriangle$. The curves give the corresponding theoretical predictions calculated from Eq.~(\ref{eq:lzmap}) for $s_3=\pm1$ (dotted curves), $s_3=\pm 0.87$ (dashed curves), $s_3=\pm 0.50$ (dot-dashed curves), and $s_3=0$ (solid curve).}
\end{figure}
\begin{figure}
\includegraphics[scale=0.89]{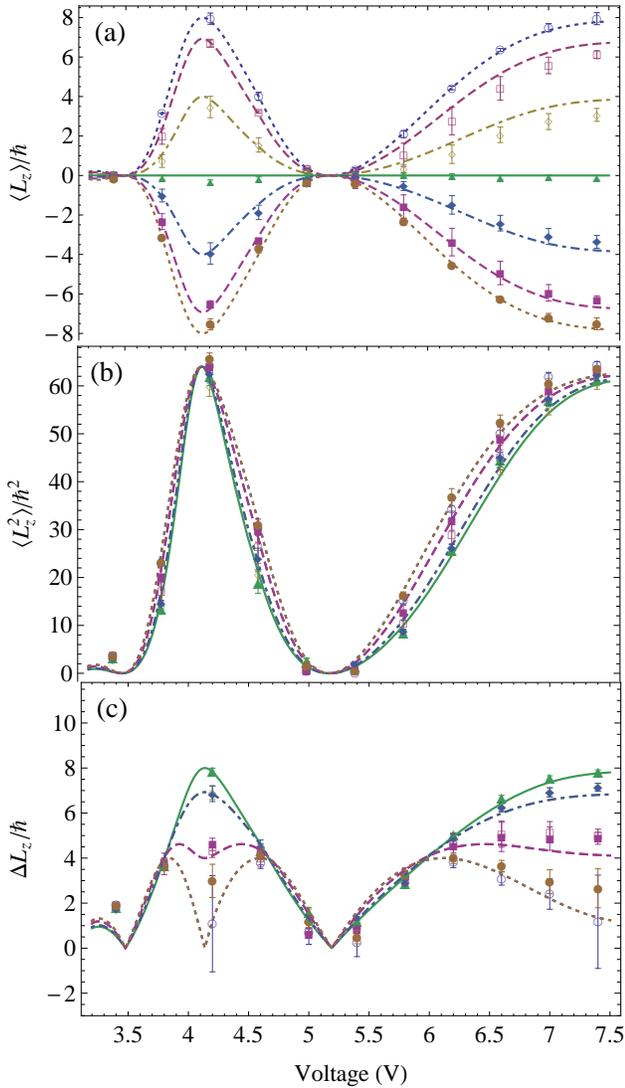}
\caption{\label{fig:fig3} Behaviour of the first (a) and second (b) moment and of the root mean square deviation (c) of the OAM distribution of the beam as a function of the voltage applied to the $q$-plate to change $\theta$. All symbols are defined just as in Fig.\ \protect\ref{fig:fig2} except for the q-plate charge, which is $q=4$, thus giving rise to superpositions of OAM states with $l=0$ and $l=\pm2q=\pm8$.}
\end{figure}

\textit{Experimental demonstration}. In order to validate our method, after calibration on OAM eigenstates, we measured the moments $\langle L_z \rangle$ and  $\langle L_z^2 \rangle$ of light beams containing different OAM superpositions. These beams were all obtained from a Gaussian TEM$_{00}$ laser beam using $q$-plates (QP) with different charges $q$, while varying the following parameters: (i) the polarization before the QP, (ii) the QP birefringent retardation (i.e., the QP ``tuning''), and (iii) the relative alignment of the beam axis and the QP center. After passing through the QP, the beam polarization is reset to the 45$^\circ$ linear polarization needed for the subsequent OAM measurement.

When the QP is centered on the beam axis, for each $q$ the generated OAM superpositions involve just three OAM eigenstates, i.e.\ $l=0,\pm 2q$. The relative weight of these three states in the superposition can be controlled using the polarization input helicity $s_3$ before the QP and the QP phase retardation $\theta$, the latter being controlled by the QP voltage $V$. The theoretical expressions for the first and second moments of the resulting OAM distribution can be easily calculated and are given by
\begin{equation}\label{eq:lzmap}
\langle L_z \rangle = \frac{2 q s_3 \sin^2{\theta/2}}{1-(1-s_3^2)\cos^2{\theta/2}}
\end{equation}
and
\begin{equation}\label{eq:lz2map}
\langle L_z ^2\rangle = \frac{4 q^2 \sin^2{\theta/2}}{1-(1-s_3^2)\cos^2{\theta/2}}.
\end{equation}
The (device-dependent) tuning-function $\theta(V)$ is obtained from an independent characterization of the QP birefringent retardation \cite{piccirillo10}. The results of our measurements on these OAM superposition states are shown in Figs.~\ref{fig:fig2} and \ref{fig:fig3}, together with the theoretical predictions from Eq.~(\ref{eq:lzmap}) and (\ref{eq:lz2map}). The agreement is clearly very good, particularly considering that there are no adjustable parameters in the theory. However, the calibration of the setup is not entirely independent from these data, as pure OAM eigenstates correspond to the first two opposite maxima obtained in the outermost curves at a voltage of about 4.2 V (for which $\theta=\pi$ and the QP is tuned). In particular, as mentioned, the precise value of the Dove prism angle $\alpha$ with respect to the PSID plane was estimated by matching the measured values of mean OAM to the theoretical ones in these specific cases.

\begin{figure}[h!]
\includegraphics[scale=0.72]{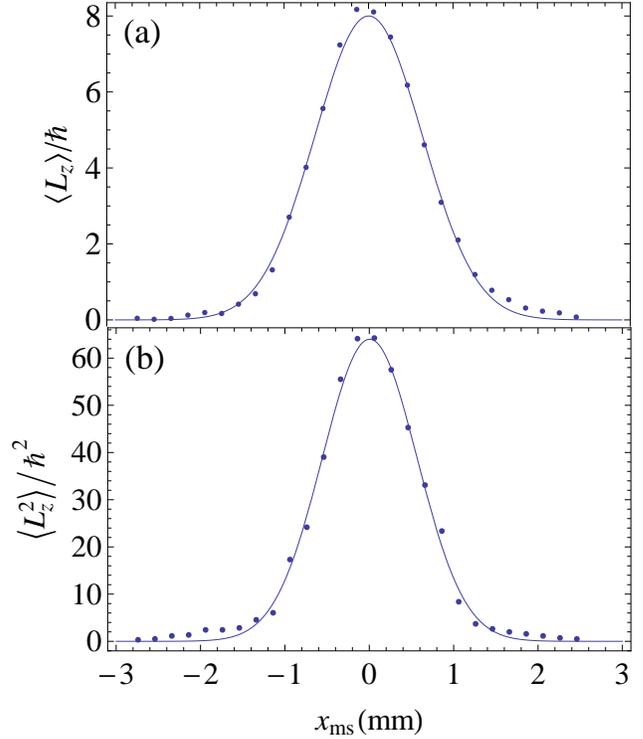}
\caption{\label{fig:fig4} Behaviour of the measured first (a) and second (b) moment of the OAM distribution of the beam as a function of the misalignment distance $x_{ms}$ between the center of the beam and the center of the vortex singularity introduced by the tuned $q$-plate with $q=4$. The theoretical predictions are represented by the solid curves. The typical experimental uncertainties on the first and second moment data-points are $\pm0.15 \hbar$ and $\pm 1.2 \hbar^2$, respectively.}
\end{figure}

In order to test the validity of our method in a more complex situation, we then measured the first and second moments of the OAM distribution of a beam generated by means of a perfectly tuned $q$-plate ($\theta = \pi$) with $q=4$, for circular polarization input ($s_3=1.0$), whose center is translated off of the beam axis by a variable distance $x_{ms}$. In this situation, the input beam in the measurement setup is similar to a Gaussian beam with an added vortex located at a distance $x_{ms}$ from the beam axis. This is not an eigenstate of OAM but is a superposition of many different values of $l$. In such case, the theoretical predictions for the first two moments of the OAM distribution can be shown to be the following: $\langle L_z \rangle = l \hbar \exp{(-x_{ms}^2/w^2)}$ and $\langle L_z \rangle = l^2 \hbar^2\exp{(-x_{ms}^2/w^2)}$, with $l = 2q$ and $w$ the beam waist in the plane of the $q$-plate. We stress that these results are based on setting the origin of the coordinate system in the beam center (and not in the vortex center). Experimentally, this is fixed by aligning the beam axis with the rotation axis of the Dove prisms in the PSID. The experimental results for this test are reported in Fig.~\ref{fig:fig4}, where they are compared with the above theoretical predictions, showing again a very satisfactory agreement.

\textit{Conclusions}. We demonstrated the possibility of directly measuring the mean and the variance of the OAM distribution of a general paraxial optical beam through a polarizing homodyne scheme based on mixing  the input beam with an azimuthally shifted copy of the same beam. The crucial role played by the actual value of the angular shift was emphasized and explicitly related to the accuracy of the measurement, the signal-to-noise ratio, the OAM bandpass and ultimately the optical-path noise within the PSID. Such a scheme could be of great help in experimental studies of OAM exchange in light-matter interaction or to weaken the usual alignment requirements typical of OAM-based free-space communication~\cite{gibson04,wang12,dambrosio12b}.

We acknowledge the financial support of the Future and Emerging Technologies (FET) programme within the Seventh Framework Programme for Research of the European Commission, under FET-Open grant number 255914- PHORBITECH.
%
\end{document}